\documentclass[12pt]{iopart}
\usepackage{graphicx}
\usepackage{bm}

\begin{document}

\title [] {Wave Functions for Fractional Chern Insulators on Disk Geometry}

\author{Ai-Lei He$^{1}$, Wei-Wei Luo$^{2}$, Yi-Fei Wang$^{1}$ and Chang-De Gong$^{1,2}$}

\address{$^1$Center for Statistical and Theoretical Condensed Matter Physics, and Department of Physics, Zhejiang Normal University, Jinhua 321004, China}
\address{$^2$National Laboratory of Solid State Microstructures and Department of Physics, Nanjing University, Nanjing 210093, China}

\ead{wangyf@gmail.com}
\vspace{10pt}
\begin{indented}
\item[]October 2015
\end{indented}

\begin{abstract}
Recently, fractional Chern insulators (FCIs), also called fractional quantum anomalous Hall (FQAH) states, have been theoretically established in lattice systems with topological flat bands. These systems exhibit similar fractionalization phenomena as the conventional fractional quantum Hall (FQH) systems. Using the mapping relationship between the FQH states and the FCI/FQAH states, we construct the many-body wave functions of the fermionic FCI/FQAH states on the disk geometry with the aid of the generalized Pauli principle (GPP) and the Jack polynomials. Compared with the ground state by exact diagonalization method, the wave-function overlap is higher than $0.97$ even when the Hilbert space dimension is as large as $3\times10^6$.  We also use the GPP and the Jack polynomials to construct edge excitations for the ferminoic FCI/FQAH states. The quasi-degeneracy sequences of fermionic FCI/FQAH systems reproduce the prediction of the chiral Luttinger liquid theory, complementing the exact diagonalization results with larger lattice sizes and more particles.
\end{abstract}

Submitted to: \emph{New J. Phys.} (Focus on Topological Physics)
\maketitle

\section{Introduction}

The integer quantum Hall (IQH) effect~\cite{Klitzing} and fractional quantum Hall (FQH) effect~\cite{Tsui} have broken the conventional framework of condensed matter theories, and have defined new topological states of matter.
In 1988, Haldane~\cite{Haldane} proposed a prototype lattice version of the IQHE without an external magnetic field, and thus without Landau levels (LLs), which has two non-trivial topological bands labeled by Chern numbers~\cite{Thouless}. The topological states in Haldane-model-like systems are also called as quantum anomalous Hall (QAH) states or Chern insulators (CIs). Based on the understanding of the CI/QAH states, and considering the strongly correlated fermionic or bosonic systems, possible FQAH states or fractional Chern insulators (FCIs) can be further conceptually defined ~\cite{Tang,checkerboard,Neupert,Sheng1,YFWang1,XLQi1,Regnault2,FCI_reviews,FCI_reviews1}. In the absence of an external magnetic field, the topological flat bands (TFBs)~\cite{Tang,checkerboard,Neupert} in CI/QAH systems play the role of LLs, and can also host the intriguing fractionalization phenomena. Since the TFBs quench the kinetic energy of particles very effectively, addition of moderate interactions push the fractionally filled CI/QAH systems into the strongly correlated fractional topological phases. Some numerical works have systematically explored the interaction effects within the TFB models, and have firmly established the existence of FCI/FQAH states~\cite{Sheng1,YFWang1,Regnault2,FCI_reviews,FCI_reviews1}.
Recently, there are various proposals of material and cold-atom realization
schemes for such FCI/FQAH systems~\cite{Xiao,Ghaemi,Lukin,Lukin1,Yannopapas,FLiu,Cooper,Cirac-TFB,Cirac-TFB1,Kapit,NYYao3}.

For the continuum FQH system of two-dimensional (2D) electron gas in a strong magnetic field, Laughlin~\cite{Laughlin}
used the lowest-Landau-level (LLL) single-particle wave functions with conserved angular momenta to construct the many-body wave function analytically for the FQH states on disk geometry. Then some other works pointed out the wave functions can also be constructed for continuous systems on different geometries like a sphere, a torus or a cylinder~\cite{Haldane1,Haldane2}. Later works found that the Laughlin wave function can be decomposed with the help of the Jack symmetric polynomial (Jacks)~\cite{Jacks,Jacks1}, which naturally reflects the generalized Pauli principle (GPP)~\cite{GPP1,GPP2,GPP3,GPP4} in the Fock space for both the Abelian and the non-Abelian FQH states. Just as the Pauli exclusion principle corresponds to the Fermi statistics, the GPP corresponds to the fractional statistics of anyons.

Recently, finding the suitable trial wave functions of FCI/FQAH states becomes an outstanding problem. It is conjectured that there is a one-to-one mapping between IQH and CI/QAH states~\cite{XLQi1,XLQi2,XLQi3,XLQi4,XLQi5}. In other words, single-particle orbitals of an LL can be mapped to the ones in a TFB. Based on the one-dimensional (1D)
maximally localized Wannier functions, Qi et al have used a mapping between the FQH and the FCI/FQAH states to construct generic wave functions of FCI/FQAH states on cylinder geometry with these 1D Wannier functions~\cite{XLQi1,XLQi2,XLQi3,XLQi4,XLQi5}. In terms of this mapping relationship, the Haldane pseudo-potential~\cite{Haldane1} for these FCI/FQAH states can be constructed~\cite{HPP1,HPP2} through a proper gauge choice for Wannier functions.
An improved prescription has been adopted to construct variational wave functions of FCI/FQAH states on torus geometry by utilizing the gauge-fixed (non-maximally) localized Wannier states~\cite{YLWu,YLWu1,Scaffidi}. From another aspect, conventional FQH states can also be obtained for 2D lattices analytically by using the conformal field theory~\cite{Nielsen,Nielsen1,Nielsen2}.
In contrast to the above analytical or semi-analytical approaches, we here pursue a very direct yet effective purely numerical prescription to construct FCI/FQAH states on disk geometry, without any variational parameter or adjustable gauge freedom, but just utilizing the powerful GPP and the Jacks structure of FQH states, and the information from exact numerical single-particle orbitals of TFB models.

In this paper, we exploit the single-particle wave functions of CI/QAH states (in the Kagom\'{e} model~\cite{Tang,WWLuo} and the Haldane~\cite{Haldane,YFWang1} model) with TFB parameters on disk geometry, and explore the polynomial structure of the continuum Laughlin wave functions~\cite{Laughlin} to establish the many-body wave functions of the FCI/FQAH states. We firstly construct the finite-size lattices on disk geometry and diagonalize the non-interacting Hamiltonian matrix to obtain the single-particle CI/QAH wave functions, then use the GPP~\cite{Haldane1} and the Jacks~\cite{Jacks,Jacks} to construct the FCI/FQAH wave functions for interacting spinless fermions. In order to test the feasibility of the many-body wave functions of FCI/FQAH states, we calculate the overlaps between these wave functions (WFs) and those from ED studies.
Very high WF overlaps demonstrate the feasibility of our method even when the Hilbert space dimension is more than $3\times10^6$. Besides, our method makes it possible to treat bigger lattices filled with more fermions (the single-particle orbitals for lattice disks with 2000 sites can be easily constructed, and the Jacks can provide the basis states for systems with more than 15 particles). We also construct the edge-excitation wave-functions for FCI/FQAH states on disk by using the GPP and the information of single-particle orbitals. We demonstrate that the degeneracy sequences of the edge
excitations of fermions on disk geometry are just the same as the bosonic systems~\cite{Kjall,WWLuo}, in quite good agreement with the chiral Luttinger liquid theory~\cite{XGWen}.

\begin{figure}[!htb]
  \begin{center}
  \includegraphics[scale=0.6]{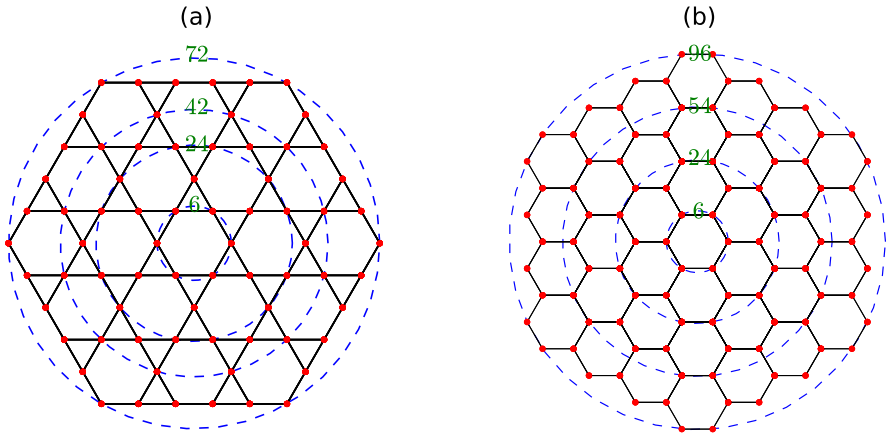}
\end{center}

\caption{(color online). (a) The kagom\'{e} lattice and (b) the honeycomb lattice on disk. The kagom\'{e} and honeycomb lattices both satisfy the $C_6$ rotational symmetry on disk. Various disk sizes are indicated by the circles and the labelled numbers.}
\label{disk_Lattice}
\end{figure}

\section{Single-particle CI/QAH orbitals on disk}

CI/QAH states are a class of topological insulators with non-trivial phases labeled by the Chern numbers~\cite{Thouless}. Haldane has proposed the first CI/QAH model~\cite{Haldane} on a honeycomb lattice in 1988. Recently, a series of CI/QAH models with TFBs were proposed~\cite{Tang,checkerboard,Neupert}. We put the CI/QAH models, e.g. the Haldane and Kagom\'{e} models, on the disks with the $C_6$ rotational symmetry (Fig. \ref{disk_Lattice}). Using tight-binding approximation and adjusting the hopping and phase parameters of CI/QAH model, we obtain the TFBs, i.e. the CI/QAH model with the specific parameters which determines a flat energy band with a high flatness ratio as well as a non-zero Chern number. An additional trap potential is required on the finite-size disk to confine the FCI/FQAH droplet and generate a soft boundary, in order that the edge modes are able to propagate around these geometries then we can observe clear edge states. We now consider the Haldane and Kagom\'{e} lattice models on disk as Ref.~\cite{WWLuo}. Here we choose the conventional harmonic trap with the form $V = V_{\rm trap}\sum_{\mathbf{r}} |\mathbf{r}|^2 n_{\mathbf{r}}$ in which $V_{\rm trap}$ is the potential strength (with the nearest-neighbor hopping $t$ as the energy unit), $|{\mathbf{r}}|$ as the radius from the disk center (with the half lattice constant $a/2$ as the length unit) and $n_{\mathbf{r}}$ as the fermion number operator ~\cite{WWLuo,Kjall}.

\begin{figure}[!htb]
  \begin{center}
\includegraphics[scale=0.55]{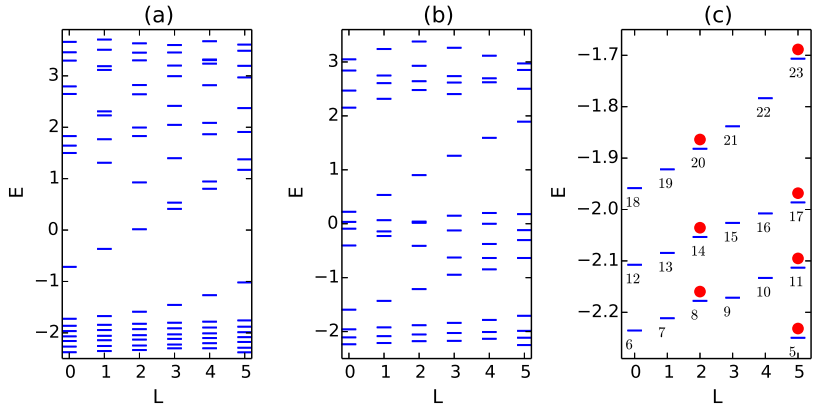}
\end{center}
  \caption{(color online). Single-particle CI/QAH orbitals on disk with TFB parameters.  Energy spectra versus the angular-momentum quantum number for (a) the honeycomb-lattice Haldane model with $N_s=96$ sites, (b) the kagom\'{e}-lattice model with $N_s=72$ sites. The harmonic trap potential parameter is $V_{\rm trap}=0.005$ for both models. The ground-state angular momentum quantum number is $5$. (c) Low-energy orbitals of the kagom\'{e}-lattice model.
  The numbers labelled under the energy levels stand for the angular-momentum quantum numbers (without modulo operation) and the orders of lowest-energy states, and the red balls represent the fermions occupying in these single-particle orbitals.}
\label{single_edge}
\end{figure}

We first notice that the Hamiltonian of the Haldane honeycomb (HC) lattice model, loaded with interacting spinless fermions~\cite{YFWang1,WWLuo}, can be written as
\begin{eqnarray}
\label{e.1}
H_{\rm HC}= &-&t^{\prime}\sum_{\langle\langle\mathbf{r}\mathbf{r}^{
\prime}\rangle\rangle}
\left[f^{\dagger}_{\mathbf{r}^{ \prime}}f_{\mathbf{r}}\exp\left(i\phi_{\mathbf{r}^{
\prime}\mathbf{r}}\right)+\textrm{H.c.}\right]\nonumber\\
&-&t\sum_{\langle\mathbf{r}\mathbf{r}^{ \prime}\rangle}
\left[f^{\dagger}_{\mathbf{r}^{\prime}}f_{\mathbf{r}}+\textrm{H.c.}\right]\nonumber -t^{\prime\prime}\sum_{\langle\langle\langle\mathbf{r}\mathbf{r}^{
\prime}\rangle\rangle\rangle}
\left[f^{\dagger}_{\mathbf{r}^{\prime}}f_{\mathbf{r}}+\textrm{H.c.}\right]\nonumber
\\&&+V_1\sum_{\langle\mathbf{r}\mathbf{r}^{ \prime}\rangle}n_{\mathbf{r}}n_{\mathbf{r}^{\prime}}
+V_2\sum_{\langle\langle\mathbf{r}\mathbf{r}^{
\prime}\rangle\rangle}n_{\mathbf{r}}n_{\mathbf{r}^{\prime}}
\end{eqnarray}
and the kagom\'{e} (KG) lattice model, also loaded with interacting spinless fermions, is written as~\cite{WWLuo},
\begin{eqnarray}
\label{e.2}
H_{\rm KG}= &-&t\sum_{\langle\mathbf{r}\mathbf{r}^{ \prime}\rangle}
\left[f^{\dagger}_{\mathbf{r}^{ \prime}}f_{\mathbf{r}}\exp\left(i\phi_{\mathbf{r}^{ \prime}\mathbf{r}}\right)+\textrm{H.c.}\right]\nonumber\\
&-&t^{\prime}\sum_{\langle\langle\mathbf{r}\mathbf{r}^{\prime}\rangle\rangle}
\left[f^{\dagger}_{\mathbf{r}^{\prime}}f_{\mathbf{r}}+\textrm{H.c.}\right]\nonumber
\\&&+V_1\sum_{\langle\mathbf{r}\mathbf{r}^{ \prime}\rangle}n_{\mathbf{r}}n_{\mathbf{r}^{\prime}}
+V_2\sum_{\langle\langle\mathbf{r}\mathbf{r}^{
\prime}\rangle\rangle}n_{\mathbf{r}}n_{\mathbf{r}^{\prime}}
\end{eqnarray}
where $f_{\mathbf{r}}$($f^{\dagger}_{\mathbf{r}}$) annihilates (creates) a spinless fermion at site $\mathbf{r}$; $\langle\dots\rangle$ , $\langle\langle\dots\rangle\rangle$ and $\langle\langle\langle\dots\rangle\rangle\rangle$ denote the nearest-neighbor (NN), the next-nearest-neighbor (NNN), and the next-next-nearest-neighbor (NNNN) pairs of sites. The phases $\phi_{\mathbf{r}^{\prime}\mathbf{r}}=\pm\phi$ which break the time reversal symmetry~\cite{Haldane}, and $V_1$, $V_2$ are the NN and NNN repulsive potentials. In order to get a TFB, we choose $t=1$, $t^{\prime}=0.60$, $t^{\prime\prime}=-0.58$,
$\phi=0.4\pi$~\cite{YFWang1} for the Haldane model and $t=1$, $t^{\prime}=-0.19$, $\phi=0.22\pi$~\cite{WWLuo} for the kagom\'{e} model.

\begin{figure}[!htb]
 \begin{center}
\includegraphics[scale=0.5]{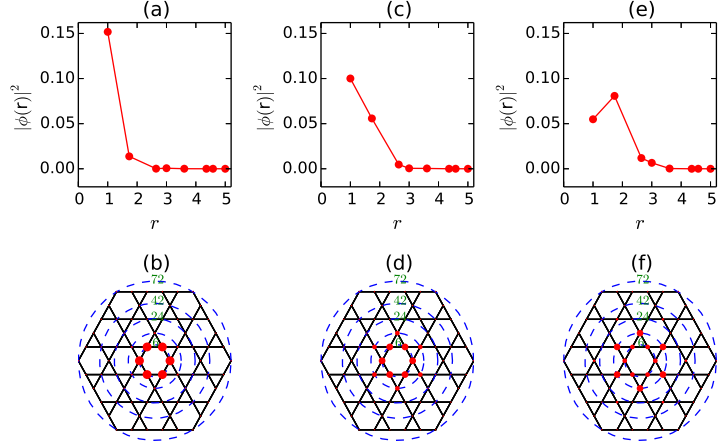}
 \end{center}
\caption{(color online). The single-particle orbitals on the kagom\'{e}-lattice disk with TFB parameters. The probability density of the single-particle orbitals on a kagom\'{e} disk with $N_s=72$ sites. The harmonic trap potential is $V_{\rm trap}=0.005$, (a-b) ground-state density profile, (c-d) the first-excited-state density profile. (e-f) the second-excited-state density profile.}
\label{disk_probality}
\end{figure}

It is easy to obtain the single-particle eigen-energies and eigen-states of the Haldane and Kagom\'{e} models on finite-size disks by constructing the lattice with the given hopping parameters and diagonalizing the non-interacting Hamiltonian matrix. In this paper, we first choose the honeycomb lattice with 96 sites and the kagom\'{e} lattice with 72 sites on disk and add a trap potential to restrain the movement of particles. The angular momentum is obviously conserved in these finite-size disk systems and these single-particle orbitals are shown in the sectors labelled by
angular momentum quantum numbers [Fig.~\ref{single_edge} (a) and (b)]. The angular momentum quantum number $L$ varies from 0 to 5 because of the $C_6$ rotational symmetry. With a harmonic trap potential $V_{\rm trap}=0.005$, the angular momentum of GS energy is 5 and the others increase with $L=4+i$ for the low energy states where $i$ denotes the $i\rm{th}$ lowest energy states [Fig.~\ref{single_edge} (c)]. By the way, we show that the GPP commands the particles occupying in these orbitals. For example, considering a $1/3$ FCI/FQAH state, there are 7 particles filled in the specific orbitals. Based on the GPP for the $1/3$ state, there is no more than one particle in three consecutive orbitals. In order to guarantee this system with the lowest energy (GS), 7 particles fill in the 19 orbitals as shown
in Fig.~\ref{single_edge} (c). At the same time, the density profile of the single-particle orbitals on lattice sites with a harmonic trap potential $V_{\rm trap}=0.005$ can be obtained along the radius of the disk as shown in Fig.~\ref{disk_probality}. The single-particle probabilities in the real space also satisfy the $C_6$ rotational symmetry(shown in Fig. \ref{disk_probality}) because of the lattice structure of the kagom\'{e}-lattice disk. With these single-particle orbitals, the next step is to generate the many-body wave functions for the FCI/FQAH states.

\section{Jack polynomial and FCI/FQAH wave functions}

The Laughlin wave function is an analytic function to describe the continuum FQH states of 2D electron gas in a magnetic field. The single-particle orbital on disk is $\phi_m(z)=(2\pi m!  2^m)^{-1/2} z^m \exp (-\vert z\vert^2/4)$ with angular-momentum quantum number $m=0,1,2,3,...$ (i.e. the angular momentum is $\rm{L}_z=m\hbar$). The $\nu=1/3$ Laughlin wave function on an infinite-size disk~\cite{Laughlin} is
\begin{eqnarray}\label{EdgeH}
\Psi_{\rm Laughlin}(\{z_i\})=\displaystyle\prod_{i<j}(z_i-z_j)^{3}\exp\left[-\displaystyle\sum_i|z_i|^2/4l_B^2\right]
\end{eqnarray}
where $z$=$x+iy$ and $l_B$ is the magnetic length. We drop the non-universal Gaussian factor ${\rm exp}({-\displaystyle\sum_i|z_i|^2/4l_B^2})$, and investigate the universal polynomial structures of $\Psi_{\rm{L}}^{1/3}(\{z_i\})=\displaystyle\prod_{i<j}(z_i-z_j)^{3}$.

As a tutorial example, we consider the $1/3$-Laughlin state with just three fermions,  and expand the wave-function $\Psi_{\rm{L}}^{1/3}(\{z_i\})$ as follows:\\
$\Psi_{\rm L}^{1/3}(z_1,z_2,z_3)=(z_1-z_2)^3(z_1-z_3)^3(z_2-z_3)^3
\\
=(+1)({z_1}^{6}{ z_2}^{3}{ z_3}^{0}-{z_1}^{6}{z_2}^{0}{z_3}^{3}-{ z_1}^{3}{ z_2}^{6}{ z_3}^{0}+{z_1}^{3}{z_2}^{0}{z_3}^{6}+{z_1}^{0}{z_2}^{6}{ z_3}^{3}-{ z_1}^{0}{ z_2}^{3}{ z_3}^{6})
\\
+(-3)({z_1}^{6}{ z_2}^{2}{ z_3}-{z_1}^{6}{ z_2}{z_3}^{2}+{ z_1}{z_2}^{6}{z_3}^{2}+{z_1}^{2}{z_2}{z_3}^{6}-{z_1}{z_2}^{2}{z_3}^{6}   -{z_1}^{2}{z_2}^{6}{z_3})
\\
+(-3)({z_1}^{5}{z_2}^{4}{z_3}^{0}-{z_1}^{5}{ z_2}^{0}{ z_3}^{4}-{ z_1}^{4}{z_2}^{5}{z_3}^{0}+{z_1}^{4}{z_2}^{0}{z_3}^{5}+{z_1}^{0}{z_2}^{5}{z_3}^{4}-{ z_1}^{0}{z_2}^{4}{ z_3}^{5} )
\\
+(+6)({z_1}^{5}{ z_2}^{3}{ z_3}-{ z_1}^{5}{ z_2}{ z_3}^{3}-{z_1}^{3}{z_2}^{5}{z_3}+{ z_1}{ z_2}^{5}{z_3}^{3}-{ z_1}{ z_2}^{3}{z_3}^{5}+{z_1}^{3}{ z_2}{z_3}^{5})
\\
+(-15)({z_1}^{4}{ z_2}^{3}{z_3}^{2}-{ z_1}^{4}{z_2}^{2}{z_3}^{3}-{z_1}^{3}{z_2}^{4}{z_3}^{2}+{z_1}^{3}{z_2}^{2}{z_3}^{4}+{z_1}^{2}{z_2}^{4}{z_3}^{3}-
{ z_1}^{2}{ z_2}^{3}{z_3}^{4}).
$\\This expansion form of the Laughlin wave function can be classified by the exponents of anti-symmetric polynomials. For example, we can write the first term $({z_1}^{6}{ z_2}^{3}{ z_3}^{0}-{z_1}^{6}{z_2}^{0}{z_3}^{3}-{ z_1}^{3}{ z_2}^{6}{ z_3}^{0}+{z_1}^{3}{z_2}^{0}{z_3}^{6}+{z_1}^{0}{z_2}^{6}{ z_3}^{3}-{ z_1}^{0}{ z_2}^{3}{ z_3}^{6})$ as $\Psi_{[6,3,0]}$. And $\Psi_{[6,3,0]}$ denotes a configuration for three particles filled in three orbitals with the angular-momentum quantum number $m=0$, $3$ and $6$ respectively. One can easily find that the $\Psi_{[6,3,0]}$ is just an (unnormalized) Slater determinant. So the $\Psi_{[6,3,0]}$ can be construed as a fermionic many-particle state in which the three particles fulfill the Pauli exclusion principle. The three-fermion Laughlin state (shown in the above) can be written as $\Psi_{\rm L}^{(3)}(z_1,z_2,z_3)=(+1)\Psi_{[6,3,0]}+(-3)\Psi_{[6,2,1]}+(-3)\Psi_{[5,4,0]}+(+6)\Psi_{[5,3,1]}+(-15)\Psi_{[4,3,2]}$. That is to say, we can decompose this three-fermion Laughlin state into five (unnormalized) Slater determinants. However, it is very difficult to expand the Laughlin states with more than five particles by such a brute-force method.

The Laughlin state can also be decomposed into symmetric monomials (for bosons) or anti-symmetric Slater determinants (for fermions) with the aid of the Jacks~\cite{Jacks,Jacks1}. The decomposition coefficients, e.g. $+1,-3,-3,+6,-15$ in the above three-fermion Laughlin state, can also be generated very effectively by the Jacks. Such an approach even works for more than 15 particles. To be explicit, we decompose the fermionic FQH states (e.g. with the filling factor $\nu=1/3$) into Slater determinants by the Jack-polynomial method~\cite{Jacks,Jacks1}:
\begin{eqnarray}
 \label{slater}
\Psi_{\rm{L}} ^{\nu}(\{z_i\})=\sum_{\mu \le \lambda} b_{\lambda \mu} sl_\mu
\end{eqnarray}
where $\lambda$ is the root partition which can index the basis state like $\lambda$=[$k0^{r-1}k0^{r-1}...k$] which shows the sequence of occupation numbers of fermions in single-particle orbitals labelled by angular-momentum quantum numbers, {$sl_\mu$} denotes a Slater determinant and $\mu$ stores the angular-momentum quantum numbers of each partition. For a particle filling number $\nu=\frac{k}{k+r}$, there is a Jacks parameter $\alpha=-\frac{k+1}{r-1}$. The expansion coefficient $b_{\lambda \mu}$ in the Eq.~(\ref{slater}) can be obtained by this recurrence relation~\cite{Jacks,Jacks1}:
\begin{eqnarray}
\label{slater2}
\ b_{\lambda \mu} = \frac{2(\frac{1}{\alpha}-1)}{\rho^{F}_\lambda (\alpha) - \rho^F_\mu (\lambda)} \sum_{\theta; \; \mu < \theta \le \lambda} (\mu_i-\mu_j) b_{\lambda \theta} (-1)^{N_{SW}}
\end{eqnarray}
where $\rho^F_\lambda (\alpha) = \sum_i \lambda_i[\lambda_i +2i (1-1/\alpha)]$. Partitions $\theta$=[$\mu_1,\mu_2,...,\mu_i +s,...,\mu_j -s,...,\mu_N$] and $\mu$=[$\mu_1,\mu_2,...,\mu_i,...,\mu_j,...,\mu_N$] are squeezed from the root partition $\lambda$. $N_{SW}$ is the number of swaps required to order the squeezed partition. We set $b_{\lambda\lambda}=1$ for simplicity.

In the following, we will show the procedure to obtain the decomposition coefficients by squeezing the partitions successively. Consider the previous $1/3$ Laughlin state with 3 fermions as an example, the Jacks parameter is $\alpha=-2$, and the root partition is $\lambda=$[$1001001$] with the angular-momentum partition [6,3,0] which satisfies the GPP. And three partitions $\lambda_1$=[$1000110$] (with angular-momentum quantum numbers [5,4,0]), $\lambda_2$=[$0110001$] (with angular-momentum quantum numbers [6,2,1]), $\lambda_3$=[$0101010$] (with angular-momentum quantum numbers [5,3,1] )are squeezed from the root partition, but preserve the original total angular-momentum quantum number. The partition $\lambda_3$ can be also squeezed from $\lambda_1$ or $\lambda_2$. The partition $\lambda_4$=[$0011100$] (with angular-momentum quantum numbers [4,3,2]) is squeezed from $\lambda_3$. The decomposition coefficients $b_{\lambda \mu}$ can be obtained recursively by the recurrence relation Eq.~(\ref{slater2}), are just the same coefficients $+1,-3,-3,+6,-15$ which have been obtained by the previous brute-force method. And the $1/3$ Laughlin state can be written by the Jacks in the second-quantization form as $\Psi_{\rm L}^{1/3}(\{z_i\})={\rm J}_{1}|{1001001}\rangle-{\rm J}_{2}|{1000110}\rangle-{\rm J}_{3}|{0110001}\rangle+{\rm J}_{4}|{0101010}\rangle-{\rm J}_{5}|{0011100}\rangle$ where ${\rm J}_{i}$ is related to the decomposition coefficient $+1,-3,-3,+6,-15$ while now being normalized. The $|{1001...}\rangle$ corresponds to the normalized Slater-determinant state for fermions. Since there are five different partitions (i.e. five Slater-determinant states), and thus the squeezed Hilbert-space dimension ${\cal D}_{\rm Jacks}=5$ for this $1/3$ Laughlin state with three fermions.

Now consider the mapping between an LL and a TFB orbital on disk geometry, and we also assume that the many-body states of FCI/FQAH have similar polynomial structure as the continuum FQH state on disk. We can construct the many-body wave functions with the Jacks structure of the Laughlin FQH states while using the TFB single-particle orbitals (instead of the original LL orbitals). We have mentioned before that, the single-particle state in an LL is of the form $\phi_{m}(z_i)\sim z^m_i$. The above Slater-determinant states, e.g. $|{1001001}\rangle$, with an angular-momentum partition [6,3,0], can be written as an unnormalized form,
$|{1001001}\rangle \sim \Psi_{[6,3,0]} \sim
 [\phi_6{(z_1)}\phi_3{(z_2)}\phi_0{(z_3)}-\phi_6({z_1})\phi_0({z_2})\phi_3({z_3})-
 \phi_3({z_1})\phi_6({z_2})\phi_0({z_3})+\phi_3({z_1})\phi_0({z_2})\phi_6({z_3})
 +\phi_0({z_1})\phi_6({z_2})\phi_3({z_3})-\phi_0({z_1})\phi_3({z_2})\phi_6({z_3})]$.
Now based upon the one-to-one correspondence between LL and TFB orbitals (with the increasing sequence of angular-momentum quantum numbers), we can replace $\phi_{m}(z_i)$ in the Slater determinants by the numerically obtained CI/QAH single-particle orbitals with the TFB parameters [Fig. \ref{single_edge}(c)], then the Jack polynomials of the conventional FQH states can be translated into the many-body FCI/FQAH lattice states. Such a numerically obtained FCI/FQAH lattice state is labelled as $\Psi_{\rm Jacks}$ since it carries the same Jacks structure as the conventional FQH states. Just to keep things straight, there are 7 particles filled in the TFB single-particle orbitals in Fig. \ref{single_edge}(c) and we begin with this configuration for the FCI/FQAH state with the minimum angular momentum for $1/3$ Laughlin filled based on the GPP. This configuration can be recorded as $|{1001001001001001001}\rangle$ and it is the root configuration that the others can be generated from it with the fixed same angular momentum using the above mentioned squeezing rules. The edge excitations of the Laughlin state on disk can be generated by the symmetric polynomials $s_n=\sum_i {z_i}^n $ ~\cite{XGWen}. The root configurations for the first excited states can be written as $|{10010010010010010001}\rangle$. Based on the above procedures, the many-body lattice wave functions for the $\nu=1/3$ FCI/FQAH states can be constructed numerically.

\begin{figure}[!htb]
 \begin{center}
\includegraphics[scale=0.5]{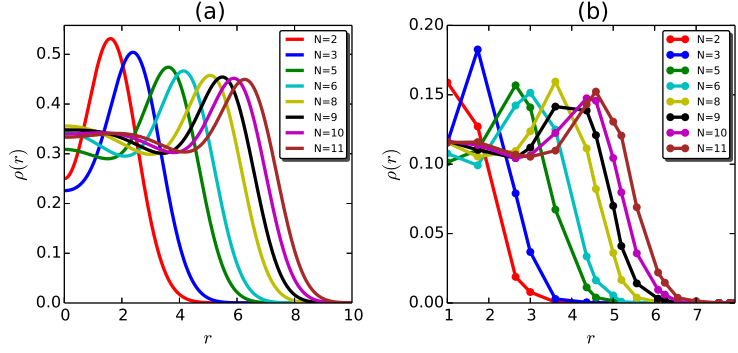}
 \end{center}
  \caption{(color online). (a) The real-space density profile of the continuum $1/3$ Laughlin state on disk. With the increasing of particles, $\rho(0)\to 1/3$. (b) The real-space density profile of the constructed $1/3$ FCI/FQAH states $\Psi_{\rm Jacks}$  on the 180-site kagom\'{e}-lattice disk.}
\label{Density_Profile_of_Laughlin}
\end{figure}

The single-particle density operator is defined as:
\begin{eqnarray}
\label{single_density}
\ \hat{\rho}(\bm{r}) = \sum_{i=1}^N\delta(\bm{r}-\bm{r}_i)
\end{eqnarray}
where {$\bm{r}_i$} denotes the position in a 2D coordinate, $N$ is the particle number, and $\delta(\bm{r}-\bm{r}_i)$ is a 2D delta-function. For a given wave function $|\Psi(\bm{r}_1,\bm{r}_2,\ldots,\bm{r}_N )\rangle$, the density function is defined as
\begin{eqnarray}
\label{single_density_1}
\rho(\bm{r})=\frac{\langle\Psi(\bm{r}_1,\bm{r}_2,\ldots,\bm{r}_N ) |\hat{\rho}(\bm{r})|\Psi(\bm{r}_1,\bm{r}_2,\ldots,\bm{r}_N )\rangle}{\langle\Psi(\bm{r}_1,\bm{r}_2,\ldots,\bm{r}_N ) |\Psi(\bm{r}_1,\bm{r}_2,\ldots,\bm{r}_N )\rangle}
\end{eqnarray}

It is a difficult task to calculate $\langle\Psi(\bm{r}_1,\bm{r}_2,\ldots,\bm{r}_N ) |\hat{\rho}(\bm{r})|\Psi(\bm{r}_1,\bm{r}_2,\ldots,\bm{r}_N )\rangle$ for a many-particle system analytically. The Jacks provides a shortcut to obtain the occupation density. With the expansion coefficient $b_{\lambda \mu}$ of every partition, it's easy to obtain the occupation probability $\rho_m$ in a specific orbital with the angular momentum quantum number $m$ on disk or sphere geometries. The real-space density $\rho(r)$ of the Laughlin state on disk can be written with the aid of the Jacks:
\begin{eqnarray}
\label{density}
\rho(\bm{r}) = \sum_{m} \frac{1}{m!2^m} \bm{r}^{2m}\exp(-\bm{r}^2/2)~\rho_m
\end{eqnarray}
where $\rho_m$ denotes the occupation density in the $m$th orbital and the factor $\frac{1}{m!2^m} \bm{r}^{2m}\exp(-\bm{r}^2/2)$ is related to the single-particle orbitals of the 2D electron gas in a strong magnetic field. The real-space density profile of the $1/3$ Laughlin GS on disk is shown in Fig.~\ref{Density_Profile_of_Laughlin} (a), with the increase of the particle number, $\rho(0)\to \frac{1}{3}$. Inspired by this, the real-space density profile of the FCI/FQAH states on disk can be obtained [shown in Fig. \ref{Density_Profile_of_Laughlin}(b)].  Here we
choose the 180-site kagom\'{e}-lattice TFB model with the $\frac{1}{3}$ filling and a harmonic trap potential $V_{\rm trap}=0.005$ along the radial direction. The density profile of the FCI/FQAH states exhibit similar structures as the Laughlin states,and also similar variation tendency when the number of fermions is increased. We should note that the particle density in the FCI bulk (i.e. the average number of fermions per lattice site) is around $0.11$ as shown in Fig. \ref{Density_Profile_of_Laughlin}(b), since the filling number $1/3$ for the FCI systems is for the filling of fermions in orbitals but not in lattice sites. The kagom\'{e} lattice has three energy bands (due to a three-site unit cell), fermions here only effectively occupy the single-particle orbitals of the lowest energy band (i.e. the TFB), and thus the particle density in the bulk of the $1/3$ FCI states should be close to the value $(1/3)\times(1/3) =1/9$.

\begin{figure}[!htb]
 \begin{center}
\includegraphics[scale=0.7]{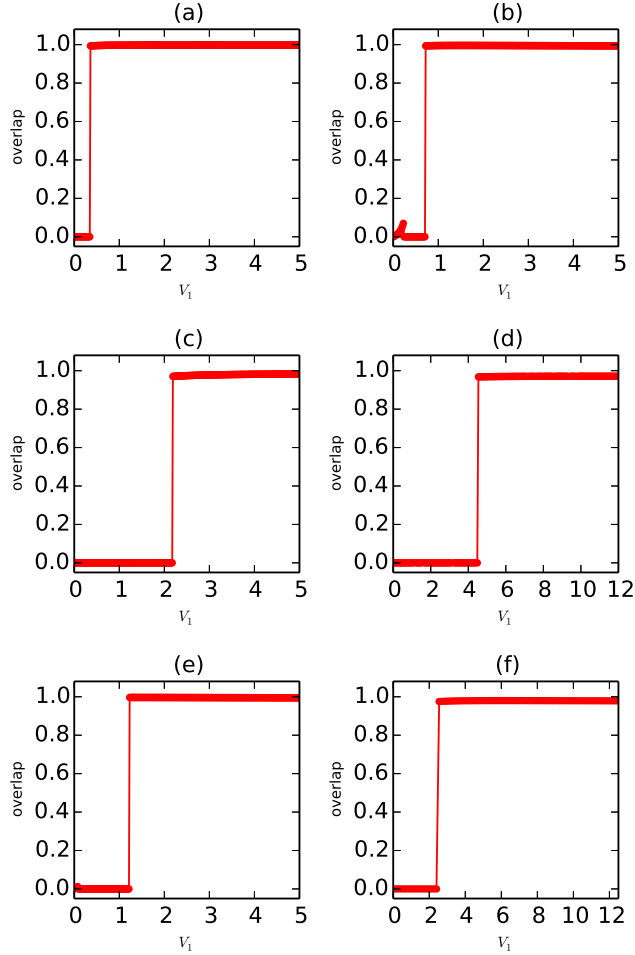}
 \end{center}
  \caption{(color online). Comparison of wave functions constructed by the Jacks with that of the ED studies, e.g. ${\cal O}=|\langle \Psi_{\rm{ED}}|\Psi_{\rm{Jacks}}\rangle|$. In these figures, the red solid circle represents the overlap between the wave functions of $1/3$ FCI/FQAH states constructed by the Jacks with the ED results for different repulsion interactions $V_1$. (a)-(c) The GS of the 42-site kagom\'{e}-lattice disk filled with 2,3 and 4 fermions, (d) The GS of the 54-site kagom\'{e}-lattice disk filled with 5 fermions. (e)-(f) The first excited states of the 54-site kagom\'{e}-lattice disk filled with 3 and 4 fermions. The harmonic trap potential is set as $V_{\rm trap}=0.005$ as before.}
\label{overlap}
\end{figure}

\section{Comparison of the wave functions}
It is almost impossible to write down an exact wave function of FCI/FQAH states analytically like the Laughlin wave functions because of the discontinuity of the CI/QAH lattice systems in contrast to the continuity of 2D electron gas. Though we cannot describe FCI/FQAH states by an analytical expression exactly, the wave functions of the FCI/FQAH states can be written as $\Psi_{\rm{L}}^{1/3}(\{z_i\})=\displaystyle\sum_i{\rm{J}}_{i}|1..1..1...\rangle_{i}$ by the Jacks~\cite{Jacks} where ${\rm{J}}_i$ are normalized Jacks coefficients of the corresponding FQH state and $|1..1..1...\rangle_{i}$ denotes the Slater determinants. Instead of the IQH state which obeys the Pauli exclusion principle, the particles in the FQH system obey the GPP. From the ED studies on the torus geometry, we know that it is necessary to add the interaction potential $V_1$ on the NN bonds to obtain the $1/3$ FQH phase. In order to estimate the rationality of the many-body wave functions constructed by using the Jacks, we can compare it with the ED results of wave functions in the real space, e.g. calculating ${\cal O}=|\langle \Psi_{\rm{ED}}|\Psi_{\rm{Jacks}}\rangle|$ for FCI/FQAH states. We first consider the FCI GS wave function of the kagom\'{e} lattice and the results of the WF overlaps are shown in Fig. \ref{overlap} (a)-(d). For the large enough interaction potential, we can observe a high overlap. Though the Jacks provides a much fewer squeezed Hilbert-space dimension than the ED dimension (show in Table. \ref{overlap_max}), the values of the overlaps for the kagom\'{e} lattice are quite high. It is worth emphasizing that the maximum value of overlaps between the ED WFs and the Jacks WFs is 0.998, 0.996, 0.984, 0.971 for the systems with $N_f=2$, $3$, $4$ and $5$ fermions. We find that a good WF overlap might be also observed even without the trap potential. When the trap potential parameter $V_{\rm trap}$ varies from $0.0$ to $0.0065$ for the above cases, the overlap becomes increasingly higher. However when the trap exceeds $0.0065$, the overlap decreases, and the best WF overlap needs an intermediate trap potential.

\begin{table}
\caption{The overlaps, ${\cal O}=|\langle \Psi_{\rm{ED}}|\Psi_{\rm{Jacks}}\rangle|$, between the $1/3$ FCI/FQAH states constructed from the Jacks and that of the ED calculations for the kagom\'{e}-disk. The overlaps for the ground states (${\cal O}_{\rm GS}$) and the first excited states (${\cal O}_{\rm ES}$) with various fermion numbers ($N_f$'s) and lattices sites ($N_L$'s) are listed. The Hilbert-space dimensions of the ED calculations (${\cal D}_{\rm ED}$) and that of the Jacks (${\cal D}_{\rm Jacks,GS}$ for the ground-state Jacks, ${\cal D}_{\rm Jacks,ES}$ for the first-excited-state Jacks) are also listed for comparison.}
\footnotesize
  \begin{center}
\begin{tabular}{c c c c c c c c c}
\br
~$N_f$~ & ~$N_L$~ & ~${\cal D}_{\rm ED,{GS} }$~ & ~${\cal D}_{\rm Jacks,GS}$~ & ~${\cal O}_{\rm GS}$~&~$N_L$~&~${\cal D}_{\rm ED,ES}$~ & ~${\cal D}_{\rm Jacks,ES}$~ & ~${\cal O}_{\rm ES}$~\\
\mr
2& 42 & 861 & 2 & 0.998 &42 &  861 & 2 & 0.999\\
3& 42 & 11480 & 5 & 0.995 &42 &  11480& 6 & 0.996\\
4&42 & 111930 & 16 & 0.984 &42 &  111930& 20 & 0.980\\
5 &54 & 3162510 & 59 & 0.971 &66 &8936928& 77 & 0.903 \\
 \br
\end{tabular}\\
  \end{center}

\label{overlap_max}
\end{table}
\normalsize

With the help of the GPP, the excited states can also be constructed. It is very easy to obtain the first excited states in the FCI/FQAH phases like that of the GS. Here we show the overlap (Fig.~\ref{overlap} (e) and (f)) between the exact excited state wave function of kagom\'{e}-lattice FCI/FQAH state and the constructed Jacks states. The maximum value of the WF overlaps is 0.999, 0.996, 0.980, 0.903 for 2-, 3-, 4- and 5-fermion systems respectively (show in Table. \ref{overlap_max}). By the way, the overlaps of the kagom\'{e}-lattice excited states are also very robust when varying the
repulsion interaction $V_1$.

\section{Edge excitations on disk}

Edge excitations can be exploited to extract the crucial information of topological order of some continuum FQH liquids~\cite{XGWen,Wan-Yang,Wan-Yang1}. In a previous work~\cite{WWLuo}, we investigated the edge excitations of the bosonic FCI/FQAH phases~\cite{YFWang1} on a disk geometry using systematic numerical exact diagonalization (ED) studies and obtained the edge excitation spectra which is quite coincident with the chiral Luttinger liquid theory~\cite{XGWen}. ED provides an accurate tool to deal with FCI/FQAH edge excitations, however it is difficult to study a large lattice system with more than a hundred sites and/or with more than several particles due to the Hilbert space limitation. The edge excitations of the Laughlin state on disk can be generated by the symmetric polynomials $s_n=\sum_i {z_i}^n $ ~\cite{XGWen}, and recently the Jacks has been employed to construct the edge states for the continuous FQH systems~\cite{Lee-Wan}.

\begin{figure}[!htb]
 \begin{center}
\includegraphics[scale=0.55]{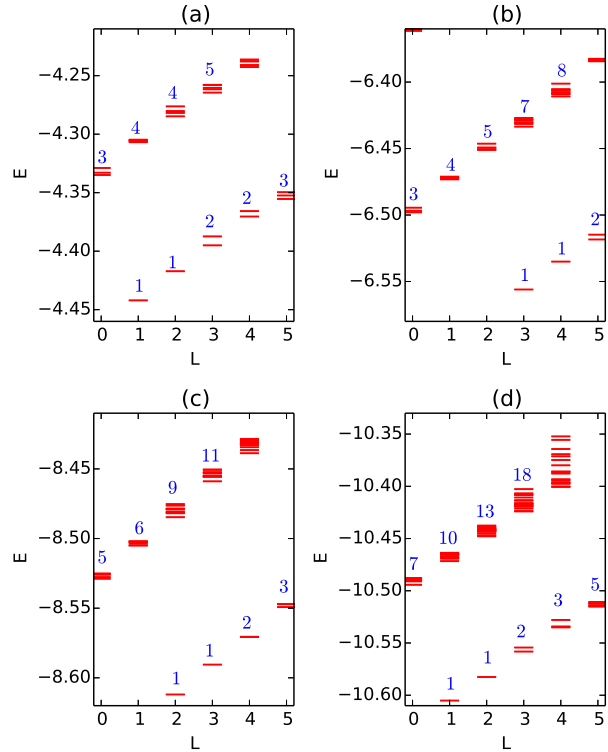}
 \end{center}
  \caption{(color online). Edge excitations of the FCI/FQAH states on the $90$-site kagom\'{e}-lattice disk filled with various fermions, $V_{\rm trap}=0.005$. The labeled numbers show the quasi-degeneracy of the low-energy excitations. (a)-(d) show us the quasi-degenerate sequences for 2, 3, 4 and 5 fermions respectively}.
\label{HC_edge}
\end{figure}

In the above section, we have compared our FCI/FQAH wave functions with the ED results and the high overlaps verified the feasibility of constructing many-body wave functions by the GPP and the Jacks. We will here try a direct approach to estimate the relative many-body energies for the FCI/FQAH states by the Jacks and the GPP. For the $1/3$ Laughlin states, particles in the single-particle orbitals fulfill the GPP. That is to say, there is no more than one particle to occupy any three adjacent orbitals. As we all know, for the identical fermion systems, the wave functions can be generated by the Slater determinant and the approximate energy of systems can be viewed as summing all energies of every single-particle orbital. For the FQH states, the many-body states are denoted by the Jacks instead of the Slater determinant. But we can expand the Jacks based on the Slater determinants. The approximate energies of the Laughlin systems can be calculated with the aid of the GPP and the Jacks:
 \begin{eqnarray}
 E_{\rm{L}}=\sum_{i} {\rho_i E_i}
\label{density_energy}
\end{eqnarray}
where $\rho_i$ is the probability of every basis related to the coefficient of Jacks and $E_i$ is the energy of every configuration. Now, we consider the kagom\'{e}-lattice TFB model with the typical harmonic trap potential $V_{\rm{trap}}=0.005$ along the radial direction on the disk~\cite{WWLuo}. We put some particles in the single-particle orbitals (shown in Fig. \ref{single_edge}(c)). As a quick case study, we put 3 fermions in the TFB and the behavior of particles obeys the GPP. The GS energy is composed by $E_{|{1001001}\rangle},E_{|{1000110}\rangle},E_{|{0110001}\rangle},E_{|{0101010}\rangle},E_{|{0011100}\rangle}$, where $E_{|{1001001}\rangle}$ denotes the energy of the state $|{1001001}\rangle$. $E_{|{1001001}\rangle}$ is the sum of single-particle eigen-energies for fermions occupying the first, forth and seventh orbitals in Fig. \ref{single_edge}(c). Just as the excited-state WFs, the approximate excitation energies can be generated by the Jacks and the symmetric polynomials ~\cite{XGWen}. By this, we can obtain the approximate many-body energies of any FCI/FQAH state. Some representative results are shown in Fig. \ref{HC_edge}. Each state can be classified with the angular momentum quantum number $L = 0, 1, 2, 3,... \rm{mod}~6$ since the kagom\'{e} lattice has the $C_6$ rotational symmetry. The quasi-degeneracies of the edge excitations in fermionic FCI/FQAH systems are the same as that in the bosonic FCI/FQAH systems~\cite{Kjall,WWLuo}.

Estimation of the energies using the Jacks above can be viewed as estimating the kinetic energy term of the Hamiltonian, and the repulsive interaction $V_1$ term has not been taken into account explicitly. For a comparison with the ED results with finite $V_1$ values, we resort to an extrapolation scheme to effectively remove the repulsive interaction term by extrapolating the ED data in the FCI/FQAH region (with good WF overlaps and beyond some critical $V_1$ values as shown in Fig.~\ref{overlap}) to the virtual $V_1=0$ limit. An exponential fitting can be chosen to fit the variations of the ED energies with different repulsive potentials. For $42$ sites with two fermions, the absolute error of the ground-state energy is $0.000003978$. For $42$ sites with three fermions, the absolute error is $0.005233$ for the ground-state energy, and is $0.004966$ for the first-excited-state energy. For $54$ sites with three fermions, the absolute error is $0.006004$ for the ground-state energy, and is $0.006129$ for the first-excited-state energy. For $54$ sites with five fermions the absolute error of ground-state energy is $0.01024$. All of the relative errors of ground-state energies are less than $0.1\%$ for these systems.

\begin{table}
\caption{The quasi-degeneracy sequences of FCI/FQAH states on disk with various fermion numbers. This degeneracy is obtained by the GPP and the Jacks. Here, $d_n$ marks the sectors of the degenerate energy levels, $N_f$ is the number of fermions.}
\begin{center}
\begin{tabular} {c c c c c c c c c c}
 \br
     & {\bf $N_f$ =2} & { \bf $N_f$=3} & { \bf $N_f$=4} & { \bf $N_f$=5} &{\bf $N_f$=6} & {\bf $N_f$=7} & {\bf $N_f$=8} & {\bf $N_f$=9} & {\bf $N_f$=10}\\
 \mr
$d_1$ & 1 & 1 & 1 & 1 & 1 & 1 & 1 & 1 & 1 \\
$d_2$ & 1 & 1 & 1 & 1 & 1 & 1 & 1 & 1 & 1 \\
$d_3$ & 2 & 2 & 2 & 2 & 2 & 2 & 2 & 2 & 2 \\
$d_4$ & 2 & 3 & 3 & 3 & 3 & 3 & 3 & 3 & 3 \\
$d_5$ & 3 & 4 & 5 & 5 & 5 & 5 & 5 & 5 & 5 \\
$d_6$ & 3 & 5 & 6 & 7 & 7 & 7 & 7 & 7 & 7 \\
$d_7$ & 4 & 7 & 9 & 10 & 11 & 11 & 11 & 11 & 11 \\
$d_8$ & 4 & 8 & 11 & 13 & 14 & 15 & 15 & 15 & 15 \\
$d_9$ & 5 & 10 & 15 & 18 & 20 & 21 & 22 & 22 & 22  \\
$d_{10}$ & 5 & 12 & 18 & 23 & 26 & 28 & 29 & 30 & 30 \\
$d_{11}$ & 6 & 14 & 23 & 30 & 35 & 38 & 40 & 41 & 42 \\
$d_{12}$ & 6 & 16 & 27 & 37 & 44 & 49 & 52 & 54 & 55  \\
 \br
\end{tabular}
\end{center}
\label{luttinger}
\end{table}

The GPP and the Jacks provide us an alternate method to obtain the edge excitations of the FCI/FQAH states on disk geometry without resorting to the ED method. We can diagonalize a very large lattice to get the all single-particle eigen-energies. There is a recurrence relation of edge-excitation quasi-degeneracy sequence for an infinite lattice with the finite-particle counting shown in Table.~\ref{luttinger}. By utilizing this table, the degeneracy sequence can be read easily. We also can estimate the number of the particles and the size of the lattice which can generate a specific degeneracy sequence. By the way, the degeneracy is not related to the types of particles, and it is only controlled by the GPP. Maybe for a finite-size lattice system filled with more particles, different degeneracy sectors may cross their boundaries then the degeneracy sequence will be broken.

\begin{figure}[!htb]
 \begin{center}
\includegraphics[scale=0.6]{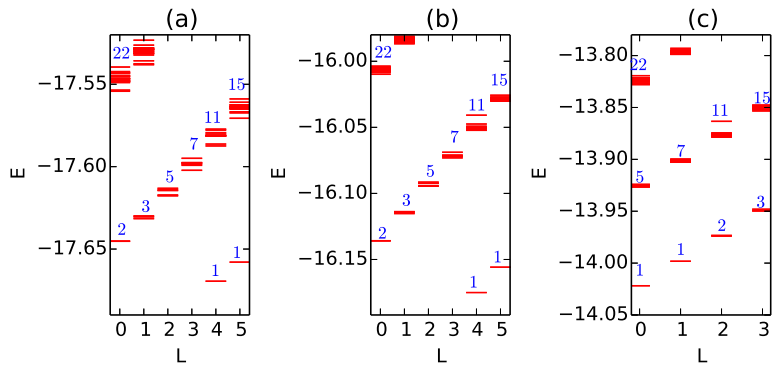}
 \end{center}
  \caption{(color online). Edge excitation in three large-size FCI/FQAH disks with $8$ fermions. The labeled numbers show the quasi-degeneracy of the low-energy excitations. (a) The honeycomb-lattice Haldane model with $228$ sites . (b) The kagom\'{e}-lattice model with $228$ sites . (c) The checkerboard-lattice model with $216$ sites. The same trap potential $V_{\rm trap}=0.005$ has been chosen for all three models.}
\label{Edge_excitation_of_CIs}
\end{figure}

There is some obvious advantage for utilizing the GPP and the Jacks to obtain the quasi-degeneracy sequence. Comparing to the costly ED method, this computation method not only is simple but also can reach very large lattice size. In Fig. \ref{Edge_excitation_of_CIs}, we show the degeneracy sequence for some over 200-site lattice systems with 8 fermions filled on disk which the ED method is not able to compute. By the way, the Haldane honeycomb-lattice model and the kagom\'{e}-lattice model own the $C_6$ rotational symmetry and the checkerboard-lattice model owns the $C_4$ rotational symmetry on disk geometry, the period of the angular momentum quantum number is 6,6,4 for the Haldane model, the kagom\'{e} model and the checkerboard model, respectively. Though they own different rotational symmetries, their degeneracy sequences are identical.

\section{Summary and discussion}
We use the numerical single-particle CI/QAH wave functions to generate the many-body wave functions for fermionic FCI/FQAH states on disk geometry. The single-particle CI/QAH eigen-states eigen-energies can be obtained by constructing a finite TFB lattice system on disk with harmonic trap potentials and diagonalizing the single-particle Hamiltonian numerically. We then utilize the GPP and the Jacks to generate the numerical many-body wave functions. Such approach yields a very direct yet effective purely numerical prescription to construct FCI/FQAH states on disk geometry, without any variational parameter or adjustable gauge freedom. In order to check the reasonability of our method, we make a WF overlap between the Jacks and the ED results. High values of the WF overlaps and the robustness against interaction confirm the rationality. The edge excitations can also be approximately obtained with the single-particle orbitals and the GPP. The quasi-degeneracy sequences of fermionic FCI/FQAH systems reproduce the prediction of the chiral Luttinger liquid theory, and they are the same as previously studied the bosonic FCI/FQAH systems, complementing the exact diagonalization results with larger lattice sizes and more particles.

We would like to mention some possible future research directions following along the present work. It is possible to further explore or utilize the FCI/FQAH wave functions at hand now, e.g., compare them with the lattice Laughlin states from the conformal field theory ~\cite{Nielsen,Nielsen1,Nielsen2} and the Gutzwiller-projected parton wave functions~\cite{parton,parton1,parton2,parton3}, and extracting the quasi-particle fractional statistics via the modular matrices~\cite{modular,Vidal,Zhu-Sheng,Zhu-Sheng1}. It would be also very interesting to extend the present approach to the non-Abelian FCI/FQAH states~\cite{YFWang2,Bernevig,Bernevig2}, the FCI/FQAH states in high-Chern-number TFBs~\cite{YFWang3,ZLiu,Sterdyniak}, and also the hierarchy FCI/FQAH states~\cite{Hierarchy,Hierarchy1}.

\section*{Acknowledgments}
We acknowledge Xin Wan for introducing the Jacks to us during the Summer School on Topological Quantum States and Phase Transitions (Jinhua, 2013). This work is supported by the NSFC of China Grants No.11374265 (Y.F.W.) and No.11274276 (C.D.G.), and the State Key Program for Basic Researches of China Grant No.2009CB929504 (C.D.G.).

\end{document}